\documentclass[runningheads]{llncs}
\usepackage{graphicx}
\usepackage{subcaption}
\usepackage{tabularx}
\usepackage{mathrsfs}
\usepackage{amsmath}
\usepackage{lineno,hyperref}
\usepackage{xcolor}
\usepackage{epsfig}
\usepackage{subcaption}
\usepackage{amssymb}
\usepackage{amstext}
\usepackage{amsmath}
\usepackage{multirow}
\usepackage{chngcntr}
\begin{document}
\title{Attention-based Interpretable Regression of Gene Expression in Histology}
\titlerunning{Interpretable Regression of Gene Expression in Histology}
\author{Mara Graziani\inst{1,2} 
\and Niccolò Marini\inst{2} 
\and Nicolas Deutschmann\inst{1} 
\and Nikita Janakarajan\inst{1,3}
\and Henning M\"uller\inst{2} 
\and Mar\'ia Rodr\'iguez Mart\'inez\inst{1}
}
%
\institute{
IBM Research Europe, 8803 R\"uschlikon, Switzerland
\and
University of Applied Sciences of Western Switzerland, 3960 Sierre, Switzerland 
\and
ETH Z\"urich, 8092 Z\"urich, Switzerland
}
\authorrunning{Mara Graziani et al.}


\toctitle{Lecture Notes in Computer Science}
\tocauthor{Authors' Instructions}
\maketitle
\begin{abstract}
Interpretability of deep learning is widely used to evaluate the reliability of medical imaging models and reduce the risks of inaccurate patient recommendations. 
For models exceeding human performance, e.g. predicting RNA structure from microscopy images, interpretable modelling can be further used to uncover highly non-trivial patterns which are otherwise imperceptible to the human eye.
We show that interpretability can reveal connections between the microscopic appearance of cancer tissue and its gene expression profiling.
While exhaustive profiling of all genes from the histology images is still challenging, we estimate the expression values of a well-known subset of genes that is indicative of cancer molecular subtype, survival, and treatment response in colorectal cancer. 
Our approach successfully identifies meaningful information from the image slides, highlighting hotspots of high gene expression. 
Our method can help characterise how gene expression shapes tissue morphology and this may be beneficial for patient stratification in the pathology unit.
The code is available on GitHub. 
\keywords{interpretability \and histopathology \and transcriptomics \and attention }
\end{abstract}
\section{Introduction}
{\let\thefootnote\relax\footnote{{\textit{Interpretability of Machine Intelligence in Medical Image Computing, MICCAI 2022}}}}
The wide variability of existing interpretability methods, in the form of post-hoc explanations~\cite{zhou2016cam,lime} or models with ad-hoc transparency constraints~\cite{rudin2019stop}, has been mainly dedicated to ensuring the safety and reliability of opaque deep neural networks.
In medical applications such as digital pathology, saliency maps highlighted the relevance of anomalous nuclei in the detection of tumorous tissue~\cite{graziani2021sharp}, and concept-based analyses confirmed that clinically relevant measures on nuclei area and appearance are learned as intermediate features~\cite{graziani2020concept}. Undesired hidden biases and behaviours were detected and corrected to improve model performance~\cite{lrphisto,make3020019,graziani2021learning}.
Only recently, interpretability techniques were proposed to uncover unknown insights about models with super-human performance, e.g. the algorithm AlphaZero~\cite{McGrath2021AcquisitionOC} defeating the world chess master.
Similarly, deep learning models can predict biomarkers invisible to the human eye~\cite{chen2020classification,schmauch2020deep}. For instance, DNA mutations~\cite{chen2020classification} and gene expression profiles were inferred from hematoxylin and eosin (H\&E) stained whole slide images (WSIs)~\cite{schmauch2020deep}, demonstrating that complex transcriptomic patterns can be captured from images without the support of sophisticated sequencing machines.

While ensuring patient safety remains the top priority in high-risk clinical environments~\cite{reviewXAIMI,lengerich2022death}, scientific discovery may benefit from unravelling the complex association between RNA expression and tissue microscopy. 
Identifying the histological patterns that are predictive of gene expression could have an important impact on diagnostic routines, facilitating early patient stratification and aiding the identification of molecular subtypes that are informative of prognosis, survival, and treatment response~\cite{guinney2015consensus}.
For example, clinical trials show that some colorectal (CRC) cancer patients with hypermutated microsatellite instability (MSI) may respond better to immunotherapy than chemotherapy~\cite{cms1immuno}.

Our central question is whether we can clarify which regions in CRC H\&E slides are the most informative about RNA transcriptomics by using a trainable attention mechanism~\cite{ilse2018attention}. 
It is yet unclear whether the existing algorithms~\cite{schmauch2020deep,weitz2021investigation} infer the bulk average expression of RNA from hotspots of high expression or as a uniform distribution on the entire slide. 
To obtain interpretable insights, we learn the expression of individual genes rather than the entire transcriptomic profile at once, training several attention-based multiple instance regression models independently.
This enables a fine grained analysis of each gene individually and reduces the need for extremely large dataset sizes~\cite{schmauch2020deep,weitz2021investigation} 
In our results, the attention mechanism brings transparency to the morphological patterns in the tissue that are learned by the model.
Hotspots of high expression are highlighted in most of our visualisations and hypotheses can be formulated to relate the histological appearance of tissue to varying gene expression. 
We show that meaningful information is successfully filtered from the WSIs by the attention, leading to more accurate gene expression estimates and patient stratification. 
A reduction of 10\% in the regression error is seen across genes and patients. 

\section{Methods}
\label{sec:methods}
\subsection{Datasets}
\label{sec:datasets}
We use images of colon adenocarcinoma (COAD) and rectal adenocarcinoma (READ) that are publicly available together with matched transcriptomic profiles at The Cancer Genome Atlas (TCGA)\footnote{\url{https://portal.gdc.cancer.gov}, as accessed in June 2022.}.
Each biopsy sample is split into three adjacent portions, two of which are used to generate H\&E-stained frozen tissue slides, called top-section (TS) and bottom-section (BS), used to verify the presence of sufficient tumour content before sequencing. The central section is used for RNA sequencing.
Differently from~\cite{schmauch2020deep,weitz2021investigation}, we focus on frozen tissue sections rather than on diagnostic slides, since they are directly adjacent to the sequenced tissue, and thus constitute the best available representation of the RNA profiles~\cite{cooper2018pancancer}. 
The WSIs are preprocessed with HistoQC~\cite{janowczyk2019histoqc} to mask out the background and blurred locations.

Gene expression profiles are obtained from the UCSC Xena Browser~\cite{goldman2020visualizing}, which links to the Genomic Data Commons~\cite{grossman2016toward} version of the TCGA COAD and READ projects.
The High Throughput-Sequencing (HT-Seq) raw counts are log2-based Fragments per Kilobase of transcript per Million mapped reads (FPKM) normalised. 
Because of a distributional shift among institutions, we retain all the patient measurements from a single institution, ignoring duplicates for 23 patients. 
We focus on the 45 genes in the ColoType signature~\cite{buechler2020colotype}, a gene set that is predictive of CRC prognosis~\cite{pan2019prognosis} and clinico-pathological variables~\cite{kheirelseid2013clinical}. The list of genes is in the Appendix Table~\ref{tab:summary-genes}.
Single-cell RNA sequencing data are also used for validation\footnote{The dataset is available at \url{https://www.weizmann.ac.il/sites/3CA/colorectal}}. 
Namely, we use the profiling of 969 single-cells from the CRC resected primary tumours of 11 patients in~\cite{li2017reference}.

As in~\cite{nguyen2021image}, we exclude patients in preoperative therapy and the rare subtypes of neuroendocrine and signet cell tumours.
In total, we use 774 WSIs at 20X magnification from 364 patients. 
The test set is built by selecting randomly 82 patients and the remaining 282 patients are used for five-fold cross validation. 
%
\subsection{Multiple Instance Regression of Gene Expression}
\label{sec:architectures-and-training}
%
\begin{figure}[ht!]
    \centering
    \includegraphics[width=\textwidth]{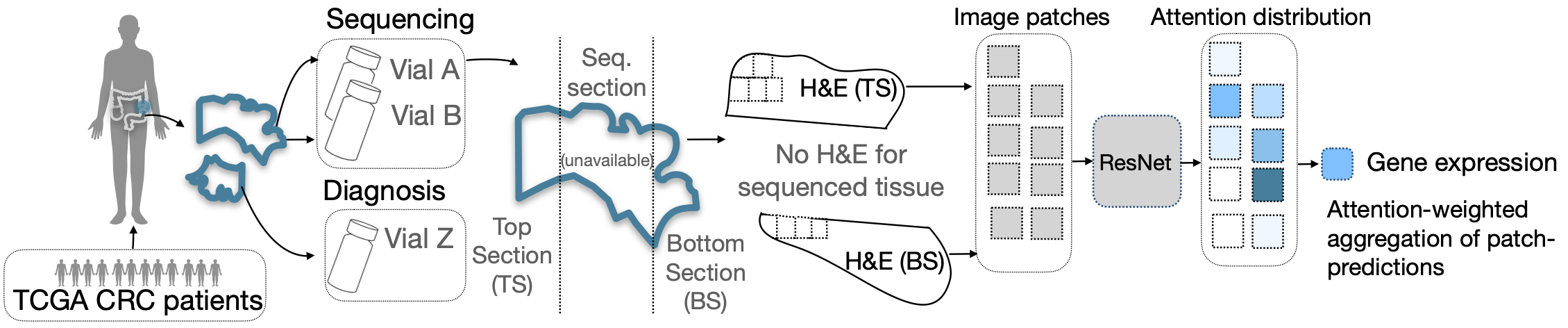}
    \caption{Attention-based Multiple Instance Learning for the semi-transparent regression of RNA gene expression from H\&E WSIs.}
    \label{fig:model}
\end{figure}
Our model is an adaptation of the attention-based multiple instance learning approach in~\cite{ilse2018attention}, which follows the implementation in~\cite{marini2022unleashing}.
We consider a patient $p$ with associated H\&E-stained scans of the top and bottom frozen tissue sections $x_t$ and $x_b$, both images in $\mathbb{R}^{w\times h}$, where $w$ is the image width and $h$ is the height. 
The associated bulk-RNA expression for a given gene $g$ is $y_g \in \mathbb{R}$. 
To deal with the large input sizes of WSIs, $x_t$ and $x_b$ are represented as a single collection of $N$ non-overlapping adjacent patches obtained by a sliding window $\{x_i\}_{i=0}^N$~\cite{marini2022multi_scale_tools}. 
The task is to learn a permutation invariant mapping to the real-valued gene label $y_g$ from this joint collection. 
We assume that each collection contains at least one instance associated with the label~\cite{Ray01multipleinstance}. 
The gene expression $y_g$ can be obtained as a linear-weighted combination of the $y_{i,g}$ predictions for each of the $x_i$ patches, where the weights are given by the trainable attention. Differently from~\cite{ilse2018attention,marini2022unleashing}, the attention weights $\{a_{i,g}\}_{i=0}^N$ are optimised to predict the continuous gene expression label rather than a binary outcome. 

The model, illustrated in Figure~\ref{fig:model}, comprises:
(i) a convolutional backbone to obtain low-dimensional representations of the input patches, i.e. a ResNet18~\cite{he2016deep} pretrained on ImageNet~\cite{imagenet};
(ii) a dense layer to predict the gene expression $y_{i,g}$ for each input $x_i$; 
(iii) an attention network to learn the attention weights. 

The output of our model is the expression of a single gene, and we train a different model for each gene in our selection. This is opposed to the approach of the SoA~\cite{schmauch2020deep}, where regression values for over 30000 gene outputs are optimised in a single training. 
To allow for a fair comparison with the SoA, we adapt their model to output the expression of a single gene at a time. The adapted model predicts the gene expression for each patch, and the predictions are then aggregated by a weighted average with larger weights given to large-valued patch predictions as in~\cite{schmauch2020deep}.
Our trainable attention removes the need for this heuristic. 
\subsection{Attention-based Model Interpretability}
\label{sec:interpretability}
The trainable attention provides transparency on how the model filters the information in the WSIs. Salient regions are identified without the need for an explicit localisation module~\cite{ilse2018attention,schlemper2019attention}.
However, recent debate argued that multiple plausible attention-based explanations may exist and that attention should be interpreted carefully~\cite{jain2019attention,wiegreffe-pinter-2019-attention}.
The research in~\cite{johnatannicolas} addressed that debate, mentioning that attention weights learned by multiple models should be ensembled by either max or average pooling to reduce the risks of obtaining misleading interpretations.
We thus average the attention weights of the models trained on the five folds, ensuring a trustworthy interpretation that focuses only on signal-bearing instances. 
\subsection{Evaluation of performance and interpretability}
\label{sec:evaluation-metrics}
Model performance is evaluated in multiple ways. 
First, the Median Average Percentage Error (MAPE) of our predictions is compared against the SoA. 
Both models are also compared to the lower bound of random guessing given by predicting the mean of the training labels. 
Finally, the significance of our results is verified by computing the non-parametric one-tailed Wilcoxon test between the patient-wise percentage error made by our model and the SoA.
Successfully regressed genes are identified as in~\cite{schmauch2020deep} by evaluating the Pearsons's correlation coefficient $\rho$ between the true and the predicted labels. 

The evaluation of model interpretability is non-trivial~\cite{weitz2021investigation,johnatannicolas}.
For morphological patterns that are known to be associated with RNA abnormal expression, e.g. mucinous tissue showing high levels of \textit{MUC2}, we verify that our explanations confirm the existing knowledge. 
For the newly identified patterns, however, there is little to no ground truth available. 
In this case, we evaluate how well patch-wise attention weights correlate to the ground truth obtained from analysing single-cell data. 
We identify, for instance, genes that are co-expressed in single-cell data according to their Pearson's correlation, and we verify that the patch-wise attentions in overlapping WSI locations correlate for those models.
Finally, the benefits of the attention are quantified by quantifying the performance improvement on an auxiliary task, i.e. inferring MSI status.  
The higher the quality of the attention-based localisation, the more accurate we expect to be the MSI-based stratification of the patients. 
\section{Experiments and Results}
 \label{sec:exp}
\subsection{Network Training}
The models are trained by optimising the mean squared error loss between the predicted gene expression value 
$\hat{y}_g$ and the label $y_g$. 
Stochastic gradient descent (SGD) is used with standard hyperparameters (learning rate 0.0001, momentum 0.9, weight decay 0.01) and early stopping (12 epochs patience).
The gradient updates affect only the dense layers and the attention mechanism. 
MSI status is learned in Sec~\ref{sec:msi-status} by two additional dense layers trained by SGD minimisation of the weighted binary cross-entropy.
For a single model trained on a GPU Tesla V100 for 5 hours, we estimate a carbon footprint of 0.65 kgCO$_2$~\cite{lacoste2019quantifying}.
\subsection{Quantitative Model Evaluation}
\label{sec:quantitative-evaluation-regression}
\begin{figure}[t!]
    \centering
    \includegraphics[width=\linewidth]{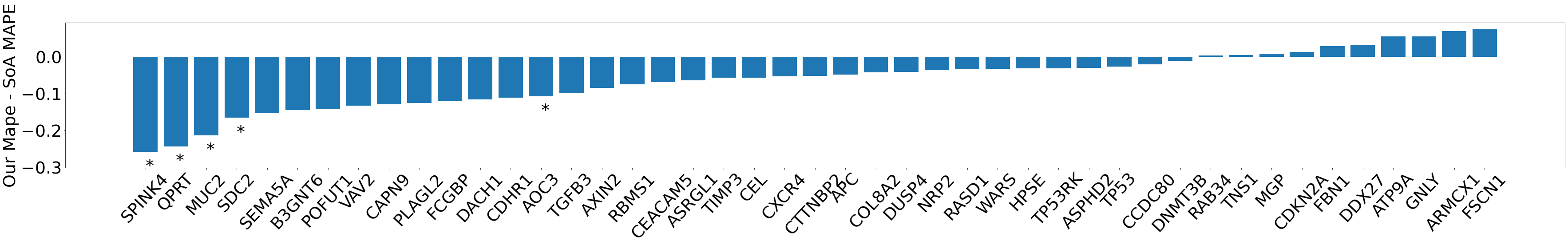}
    \caption{Median difference between the test set errors by our method and the SoA. Statistical significance is shown for p-value $<0.05$. Lower is better.}
    \label{fig:differences-ape}
\end{figure}
The gene expression estimates obtained by our method are more accurate than the SoA. Across all genes and patients, our model obtains a reduction of 10\% in the MAPE, i.e. $0.65\pm0.08$ in our model against $0.72\pm$0.12 in the SoA. Random guessing as described in Sec.~\ref{sec:evaluation-metrics} achieves MAPE $0.94\pm13.0$, showing that both methods are capturing meaningful signals from the WSIs. 
Computing the Pearson's correlation with the ground-truth labels further confirms the accuracy of our model. \textit{MGP} achieves the highest $\rho$ with 0.71 against 0.69 of the SoA (p-value$<0.0001$). 
The predictions for \textit{CCDC80}, \textit{NRP2}, and \textit{RAB34} also show strong correlation with the ground-truth, with $\rho$ $0.65$, $0.62$ and $0.61$ against the SoA $\rho$ at $0.61$, $0.61$ and $0.55$ respectively. 
The median differences in the errors made by the two models are shown in Figure~\ref{fig:differences-ape}. 
The statistical significance of the Wilcoxon test is reported for the individual patient-wise error differences. 
The detailed results on the full gene set are provided in the Appendix Figure~\ref{fig:comparison-ape} and Table~\ref{tab:best-results}.

The largest improvement is observed for \textit{SPINK4}, where the MAPE decreases to $0.62\pm1.9$ from $0.88\pm3.1$ of the SoA (p-value$<0.001$). 
\textit{QPRT}, \textit{MUC2}, \textit{AOC3} and \textit{SDC2} report significantly better MAPE than the SoA, achieving respectively $0.74\pm1.3$, $0.73\pm1.2$, $0.55 \pm 1.3$ and $0.64\pm1.4$ against $0.98\pm1.1$, $0.95\pm1.9$, $0.65 \pm 3.7$ and $0.81\pm2.3$ (p-value$<0.05$).
The Pearson's $\rho$ for \textit{QPRT} and \textit{MUC2}, in particular, increase respectively to $0.43$ and $0.46$ (p-value$<0.0001$) from $0.35$ and $0.13$ for the SoA.  
\subsection{Attention-based Identification of Hotspots and Patterns}
\label{sec:qualitative-interpretability}
\begin{figure}[!t]
    \centering
    \includegraphics[width=\textwidth]{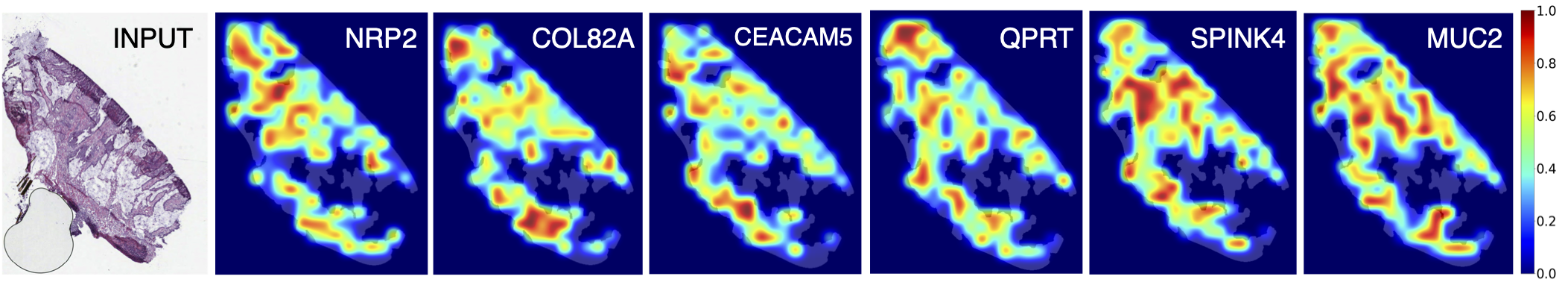}
    \caption{Localised ensemble attention on an unseen test image for six genes.}
    \label{fig:attention-maps}
\end{figure}

\begin{figure}[b]
    \centering
    \begin{subfigure}[b]{0.34\linewidth}
        \includegraphics[width=\linewidth]{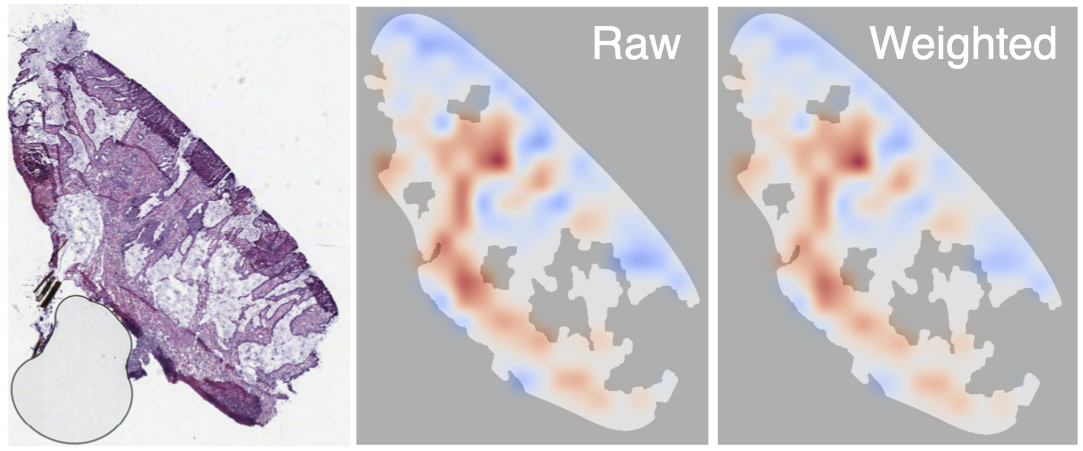}\caption{\textit{NRP2}}\label{fig:wp-nrp2}
    \end{subfigure}
    \hfill
      \begin{subfigure}[b]{0.64\linewidth}
        \includegraphics[width=\linewidth]{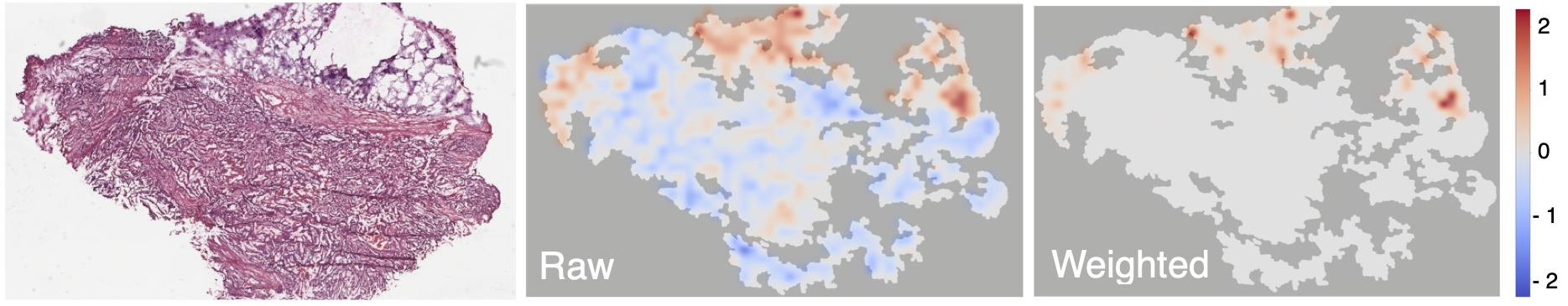}\caption{\textit{MUC2}}\label{fig:wp-muc2}
    \end{subfigure}
    \caption{Left-to-right order: Original input WSI, raw spatial predictions and attention-weighted predictions. Blue and red show low and high expression respectively.}
    \label{fig:predictions}
\end{figure}

\begin{figure}
    \centering
    \begin{subfigure}[b]{0.3254\linewidth}
        \includegraphics[width=\linewidth]{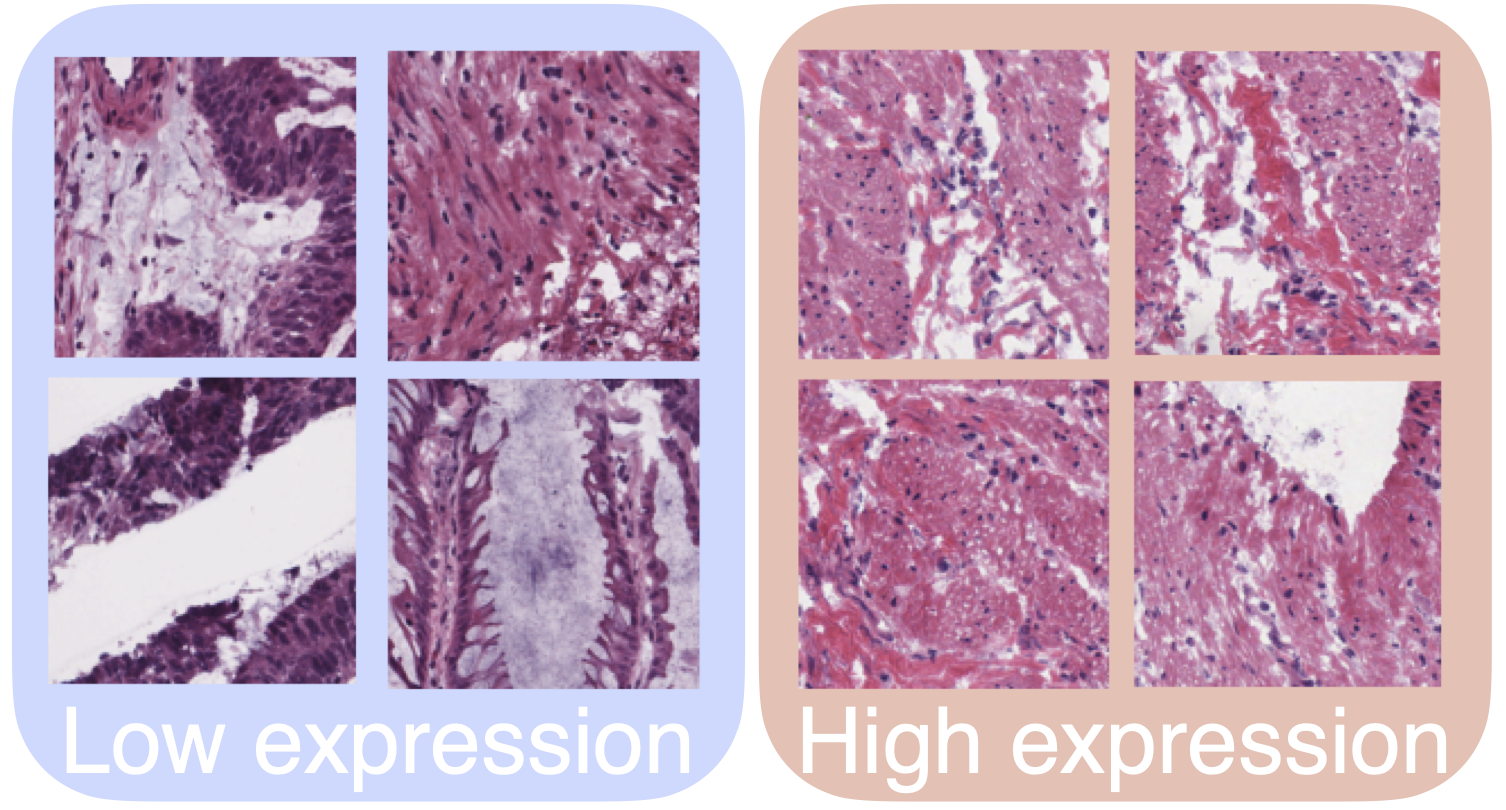}\caption{\textit{NRP2}}
    \end{subfigure}
    \hfill
      \begin{subfigure}[b]{0.3254\linewidth}
        \includegraphics[width=\linewidth]{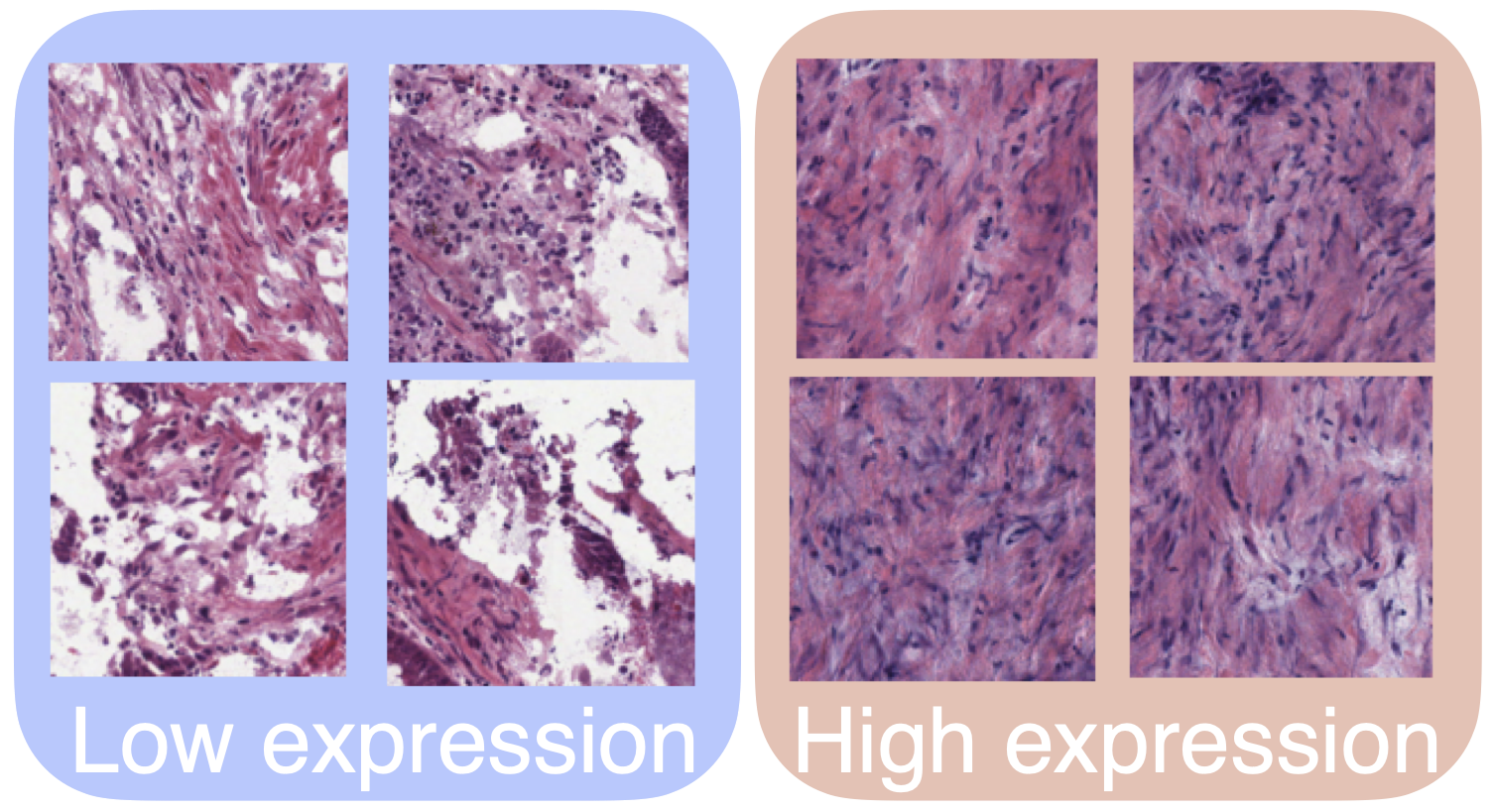}\caption{\textit{COL8A2}}
    \end{subfigure}
    \hfill
     \begin{subfigure}[b]{0.3254\linewidth}
        \includegraphics[width=\linewidth]{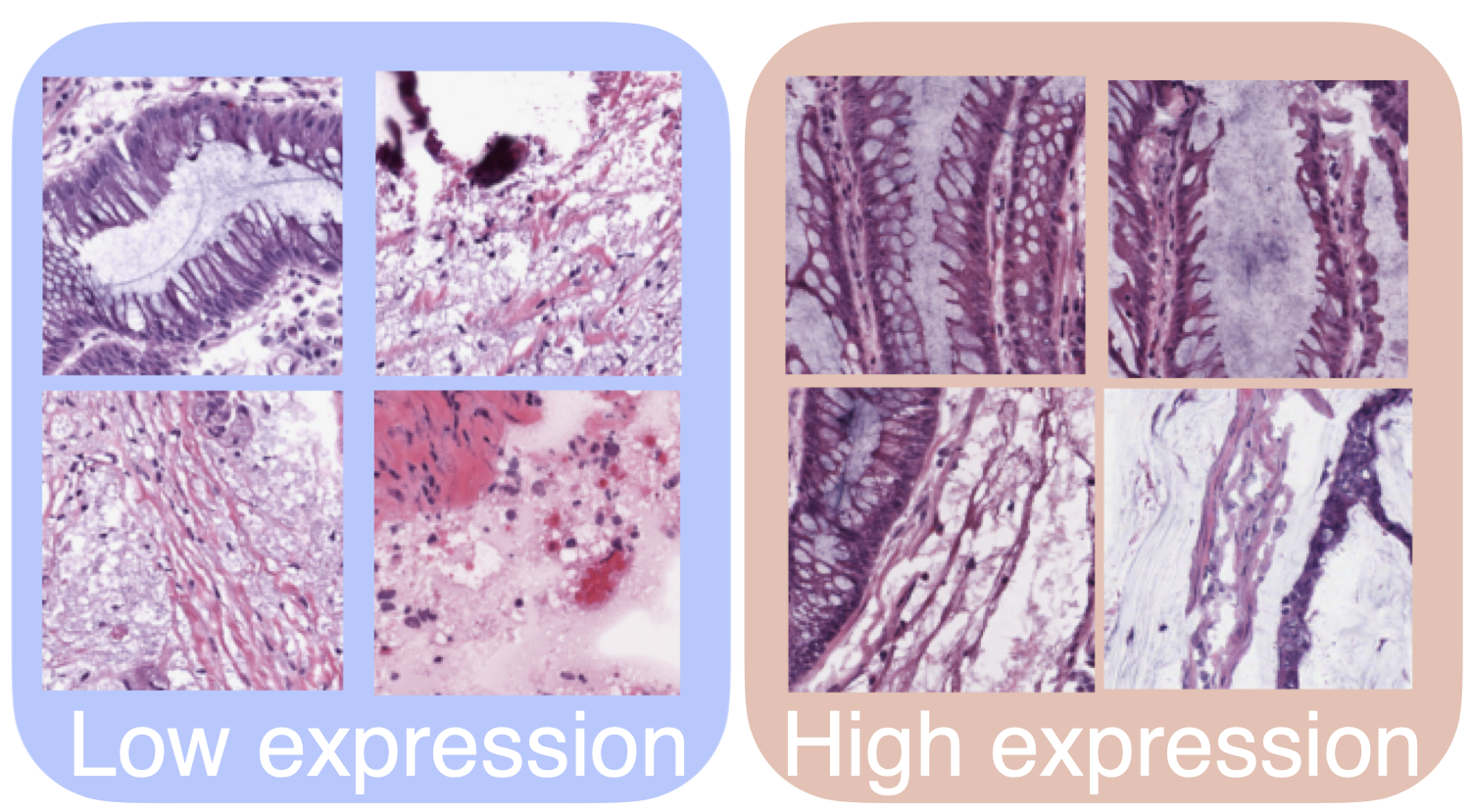}\caption{\textit{MUC2}}
    \end{subfigure}
    \caption{Patches with highest attention for predicted low (blue box) and high (red box) expression. Differing patterns are visible across genes and expression levels.}
    \label{fig:highest-att}
\end{figure}

Figures~\ref{fig:attention-maps} and~\ref{fig:predictions} show the predictions and the distribution of the attention for our best gene models on the input WSI.
The attention identifies hotspot regions, rather than being uniformly distributed over all patches. 
Figure~\ref{fig:highest-att} visualises more in detail the morphological patterns in the highlighted hotspots of high and low gene expression. 
We retrieve, for instance, the patches that received the highest normalised attention weights, namely the highest $N_pa_{i,g} \space \forall i, p$ in the testing set, where $i=1,\dots,N_p$, and $N_p$ is the total number of patches for patient $p$\footnote{The normalisation enables the comparison across patients, since $\sum_{i=0}^{N_p} a_{i,g} = 1$ }. 
Additional visualisations are in the Appendix and the code repository\footnote{}. 
%

Genes that are co-expressed in single-cell data show high attention on the same WSI areas.
For example, \textit{NRP2} and \textit{MUC2} show a correlation value of $\rho=0.40$ at the single-cell gene expression level. Similarly, they display a correlation of $0.63$ in the patch-wise attention values. 
Conversely, \textit{MUC2} is not co-expressed at the single-cell level with \textit{AOC3} and \textit{COL8A2}, and the attention localise on different regions, with $0.16$ and $0.12$ patch-wise correlations respectively. The results for all gene couples are in the Appendix Figure~\ref{fig:correlation-analysis}. 

\subsection{Quantitative Evaluation of the Attention}
\label{sec:msi-status}
The MSI status is inferred from the gene expression values regressed from the H\&E slides. The input noise is reduced by selecting only the genes that were predicted with Pearson's p-value$<0.05$ by each method. This leads to 37 genes with our method and 35 with the SoA from the initial pool of 45 genes.
We perform 5x2 cross-validation to obtain more stable estimates through repetitions on non-overlapping training sets~\cite{5x2testing}. 
We observe an increase in the AUC from $0.60\pm0.08$ of the SoA to $0.71\pm 0.05$ with our attention-based model. In comparison, using the ground-truth sequencing data yields AUC $0.86\pm0.04$. 
\section{Discussion}
\label{sec:discussions}
The experiments in Sec.~\ref{sec:quantitative-evaluation-regression} quantify the accuracy of our bulk-RNA predictions from WSIs in terms of the prediction error, i.e. MAPE, and correlation with the ground-truth labels, i.e. Pearson correlation $\rho$. 
Our trainable attention mechanism reduces the error for the majority of the analysed genes, achieving significant improvements over the SoA for the \textit{SPINK4, QPRT, MUC2, SDC2} and \textit{AOC3} genes.
Besides accuracy improvements, learning the attention removes the use of a priori heuristic choices and hyperparameter tuning in~\cite{schmauch2020deep}.

The main contribution is the analysis of the signals that are picked by the models to obtain our performance improvements. 
Focusing on one gene at a time enables a fine-grained analysis of the model attention patterns since the entire model capacity is optimised to capture salient features of individual genes. 
The attention maps in Sec.~\ref{sec:qualitative-interpretability} give interesting insights. For instance, Figure~\ref{fig:attention-maps} shows that the attention localises over hotspot regions rather than being uniformly spread across the slide.
This answers the questions raised in~\cite{weitz2021investigation} regarding the distribution of the information related to gene expression.
As expected, high attention is seen in overlapping WSI regions for genes that are co-expressed in the same cell and in differing regions for genes that are not co-expressed. 
Moreover, the highest attention weights point to visibly different histological patterns, as shown in Figure~\ref{fig:highest-att}. For example, the high-expression patches of \textit{COL8A2} present higher concentrations of stroma than the low-expression ones.

An interesting example is that of \textit{MUC2}, a particularly well-known gene that impacts tissue histology.
\textit{MUC2} over-expression leads to high amounts of visible intra- and extra-cellular mucin, characterizing the tissue as mucinous adenocarcinoma.
An example of mucinous adenocarcinoma is shown in Figure~\ref{fig:wp-muc2}. Our model predicts high values of \textit{MUC2} for the regions in the slides where the mucinous tissue is the most visible. 
When these are further weighted by the attention, the localisation is highly focused on a few hotspots of high expression. 
From a quantitative standpoint, the impact of the attention mechanism for this specific gene is remarkably significant. 
The SoA predictions for \textit{MUC2} are similar to random guessing (with MAPE $0.95$ and $\rho$ $0.13$), whereas our method significantly reduces the MAPE to $0.73$ and the correlation with ground truth values, i.e. $\rho$, to $0.46$ with p-value$<0.0001$. 
This result is also beneficial to the prediction of MSI, which is known to correlate with the presence of mucin~\cite{nguyen2021image}. 
\section{Conclusion}
\label{sec:conclusion}
We proposed an attention-based multiple instance regression model to infer bulk gene expression of tissue sections from H\&E histology slides of colorectal adenocarcinoma.
The automated regression of transcriptomics from H\&E is not yet a feasible replacement for sequencing the tissue, but the analysis of interpretability patterns can highlight some histological patterns associated with specific gene expression levels. 
We show here how the attention mechanism successfully filters informative content from the vast amount of information in the WSIs, leading to significantly lower errors and interesting insights into how gene expression impacts tissue morphology.
Further developments of this method should introduce pathologists in the loop to validate the identified patterns and further investigate their association to differentially expressed genes. 
In turn, this could help pathologists to identify cancer molecular subtypes from WSIs and consequently stratify patients for targeted treatment. 

\paragraph{Acknowledgements}
This work was supported by the Swiss National Science Foundation Sinergia project (CRSII5\_193832) and the EU H2020 project AI4Media (951911).

\bibliographystyle{unsrt}
\bibliography{omybibliography}
\newpage
\appendix
\counterwithin{figure}{section}
\counterwithin{table}{section}

\section{Description of Selected Genes}
\begin{table}[!ht]
    \centering
    \caption{Summary of the 45 genes considered in this study, including the ColoType signature and biomarkers of colorectal adenocarcinoma. }
    \label{tab:summary-genes}
    \fontsize{7}{6}\selectfont
\begin{tabularx}{\textwidth}{ccccXX}
Gene & Expression level & Associated CMS & Ref. & Description \\\hline
ASPHD2 & HIGH & CMS1 & \cite{buechler2020colotype} & Dioxygenase activity \\
ATP9A & LOW & CMS1 & \cite{buechler2020colotype} & Membran trafficking of cargo proteins \\
AXIN2 & LOW & CMS1 & \cite{buechler2020colotype} & Regulation of beta-catenin stability in WNT pathway\\
CDHR1 & LOW & CMS1 & \cite{buechler2020colotype} & Cadherin superfamily, cell adhesion\\
CTTNBP2 & LOW & CMS1 & \cite{buechler2020colotype} & Cortacting binding protein\\
DACH1 & LOW & CMS1 & \cite{buechler2020colotype} & Chromatin associated protein\\
GNLY & HIGH & CMS1 & \cite{buechler2020colotype} & Antimicrobial protein that kills intracellular pathogens.  \\
HPSE & HIGH & CMS1 & \cite{buechler2020colotype} & Enhances angiogenesis  \\
SEMA5A & LOW & CMS1 & \cite{buechler2020colotype} & Promotes angiogenesis by cell proliferation and migration and inhibites apoptosis \\
WARS & HIGH & CMS1 & \cite{buechler2020colotype} & Regulates ERK, Akt, and eNOS pathways, associated with angiogenesis\\
CEL & HIGH & CMS2 & \cite{buechler2020colotype} & \\
DDX27 & HIGH & CMS2 & \cite{buechler2020colotype} &Probable ATP-dependent RNA helicase\\
DUSP4 & LOW & CMS2 & \cite{buechler2020colotype} &Regulates mitogenic signal transduction \\
FSCN1 & LOW & CMS2 & \cite{buechler2020colotype} & Organizes filamentous actin into parallel bundles\\
LYZ & LOW & CMS2 & \cite{buechler2020colotype} \\
PLAGL2 & HIGH & CMS2 & \cite{buechler2020colotype} \\
POFUT1 & HIGH & CMS2 & \cite{buechler2020colotype} \\
QPRT & HIGH & CMS2 & \cite{buechler2020colotype} \\
TP53RK & HIGH & CMS2 & \cite{buechler2020colotype} \\
TRIB2 & LOW & CMS2 & \cite{buechler2020colotype} \\
ASRGL1 & HIGH & CMS3 & \cite{buechler2020colotype} \\
B3GNT6 & HIGH & CMS3 & \cite{buechler2020colotype} &Plays an important role in the synthesis of mucin-type O-glycans in digestive organs \\
CAPN9 & HIGH & CMS3 & \cite{buechler2020colotype} \\
FBN1 & LOW & CMS3 & \cite{buechler2020colotype} \\
FCGBP & HIGH & CMS3 & \cite{buechler2020colotype} \\
RASD1 & HIGH & CMS3 & \cite{buechler2020colotype} \\
RBMS1 & LOW & CMS3 & \cite{buechler2020colotype} \\
SPINK4 & HIGH & CMS3 & \cite{buechler2020colotype} \\
TIMP3 & LOW & CMS3 & \cite{buechler2020colotype} \\
VAV2 & LOW & CMS3 & \cite{buechler2020colotype} \\
AOC3 & HIGH & CMS4 & \cite{buechler2020colotype} &Participates in lymphocyte extravasation and recirculation \\
ARMCX1 & HIGH & CMS4 & \cite{buechler2020colotype} & Regulates mitochondrial transport during axon regeneration\\
CCDC80 & HIGH & CMS4 & \cite{buechler2020colotype} &Promotes cell adhesion and matrix assembly \\
COL8A2 & HIGH & CMS4 & \cite{buechler2020colotype} &Necessary for migration and proliferation of vascular smooth muscle cells \\
MGP & HIGH & CMS4 & \cite{buechler2020colotype} &  Thought to act as inhibitor of bone formation\\
NRP2 & HIGH & CMS4 & \cite{buechler2020colotype} & May play a role in cardiovascular development, axon guidance, and tumorigenesis\\
RAB34 & HIGH & CMS4 & \cite{buechler2020colotype} \\
SDC2 & HIGH & CMS4 & \cite{buechler2020colotype} &participates in cell proliferation, cell migration and cell-matrix interactions \\
TGFB3 & HIGH & CMS4 & \cite{buechler2020colotype} &Involved in embryogenesis and cell differentiation \\
TNS1 & HIGH & CMS4 & \cite{buechler2020colotype} \\
DNMT3B & UNK &UNK & \cite{pan2019prognosis} &Required for genome-wide de novo methylation. Seems to be involved in gene silencing\\
CDKN2A & UNK & UNK& \cite{kheirelseid2013clinical} &Many studies suggest poorer prognostic outcome for patients with hypermethylation in colorectal, liver, and younger lung cancer patients\\
CEACAM5 & UNK &UNK & \cite{kheirelseid2013clinical} & Cell adhesion, intracellular signaling and tumor progression \\
CXCR4 & UNK & UNK& \cite{kheirelseid2013clinical} & Essential for the vascularization of the gastrointestinal tract\\
MUC2 & UNK &UNK & \cite{kheirelseid2013clinical} & Coats the epithelia in the colon. May exclude bacteria from the inner mucus layer.\\

\end{tabularx}
\end{table}

\section{Detailed Model Evaluation}
\label{app:full-results}

\begin{figure}
    \centering
    \includegraphics[width=\linewidth]{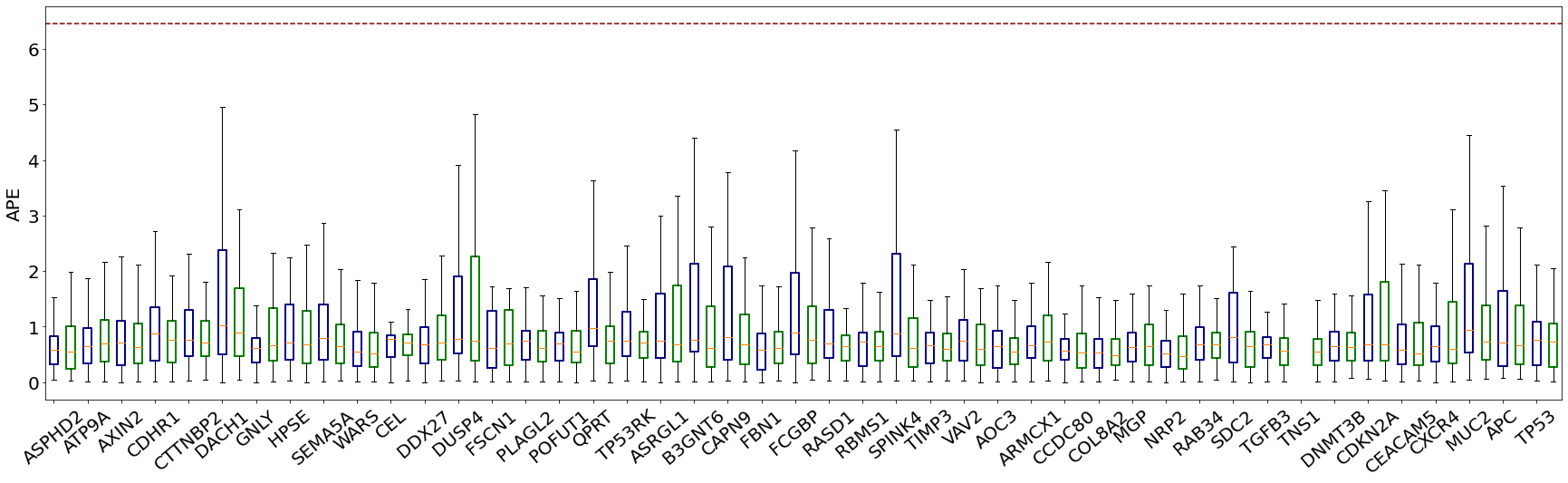}
    \caption{Absolute Percentage Error on the test set for the baseline model (in blue) and our method (in green). The models are trained on five training folds and the final prediction is obtained as the ensembled average of the predictions. The red horizontal line shows the performance of random prediction. The lower the better.}
    \label{fig:comparison-ape}
\end{figure}

Detailed results are reported in Table~\ref{tab:best-results}, which was not included in the main paper because of the limited space. The model of the gene expression for \textit{NRP2} leads to the best results overall, with MAPE=$0.47\pm1.19$ and $\rho=0.62$ against the SoA MAPE=$0.51\pm1.17$ and $\rho=0.61$ of the baseline. 

\begin{table}
 \fontsize{8}{6}\selectfont
  \caption{Comparison of the Test MAPE. Standard deviation reported in brackets.}
    \label{tab:best-results}
    \centering
    \begin{tabular}{c||c|c}
        
        \multirow{2}{*}{Gene} & \multicolumn{2}{c}{MAPE $\downarrow$} \\
         & SoA & ours\\\hline
    NRP2 & 0.51 (1.2) & 0.47 (1.2)\\
    COL8A2 & 0.54 (4.1) & 0.49 (3.5)\\
    CEACAM5 & 0.59 (0.8) & 0.52 (0.9)\\
    WARS & 0.55 (6.5) & 0.52 (12.3)\\
    CCDC80 & 0.56 (2.0) & 0.54 (1.5)\\
    TNS1 & 0.54 (1.3) & 0.54 (1.0)\\
    ASPHD2 & 0.58 (1.9) & 0.55 (2.7)\\
    AOC3 & 0.65 (3.7) & 0.55 (1.3)\\
    POFUT1 & 0.70 (0.9) & 0.55 (1.4)\\
    TGFB3 & 0.67 (6.2) & 0.57 (2.7)\\
    CXCR4 & 0.65 (3.2) & 0.60 (4.1)\\
    TIMP3 & 0.66 (3.4) & 0.60 (2.6)\\
    VAV2 & 0.74 (1.3) & 0.61 (1.3)\\
    FBN1 & 0.58 (1.6) & 0.61 (1.9)\\
    PLAGL2 & 0.74 (1.0) & 0.61 (1.4)\\
    B3GNT6 & 0.76 (5.0) & 0.62 (2.8)\\
    SPINK4 & 0.88 (3.1) & 0.62 (1.9)\\
    AXIN2 & 0.71 (0.7) & 0.63 (0.7)\\
    DNMT3B & 0.65 (9.3) & 0.64 (10.5)\\
    SDC2 & 0.81 (2.3) & 0.64 (1.4)\\
    MGP & 0.64 (1.6) & 0.65 (3.0)\\
    SEMA5A & 0.80 (1.7) & 0.65 (1.4)\\
    RBMS1 & 0.72 (1.1) & 0.65 (1.7)\\
    RASD1 & 0.69 (5.8) & 0.66 (4.5)\\
    APC & 0.71 (2.1) & 0.67 (2.0)\\
    GNLY & 0.62 (4.8) & 0.67 (3.5)\\
    ASRGL1 & 0.74 (4.1) & 0.68 (4.1)\\
    HPSE & 0.71 (6.7) & 0.68 (5.0)\\
    CDKN2A & 0.67 (36.1) & 0.69 (29.8)\\
    CAPN9 & 0.82 (5.1) & 0.69 (2.6)\\
    RAB34 & 0.69 (2.3) & 0.69 (5.5)\\
    ATP9A & 0.64 (0.9) & 0.70 (1.0)\\
    FSCN1 & 0.62 (5.0) & 0.70 (4.8)\\
    TP53RK & 0.74 (1.2) & 0.71 (0.9)\\
    CTTNBP2 & 0.76 (3.0) & 0.71 (2.2)\\
    DDX27 & 0.68 (1.0) & 0.72 (2.1)\\
    CEL & 0.77 (1.2) & 0.72 (1.4)\\
    ARMCX1 & 0.66 (15.1) & 0.73 (15.1)\\
    MUC2 & 0.95 (1.9) & 0.73 (1.2)\\
    TP53 & 0.76 (0.9) & 0.74 (0.9)\\
    DUSP4 & 0.78 (3.1) & 0.74 (3.5)\\
    QPRT & 0.98 (1.1) & 0.74 (1.3)\\
    CDHR1 & 0.87 (1.2) & 0.76 (1.2)\\
    FCGBP & 0.89 (2.5) & 0.77 (2.5)\\
    DACH1 & 1.02 (2.9) & 0.90 (2.2) \\
    \end{tabular}   
\end{table}

\section{Additional Visualizations}

\begin{figure}[h!]
    \centering
    \includegraphics[width=0.8\textwidth]{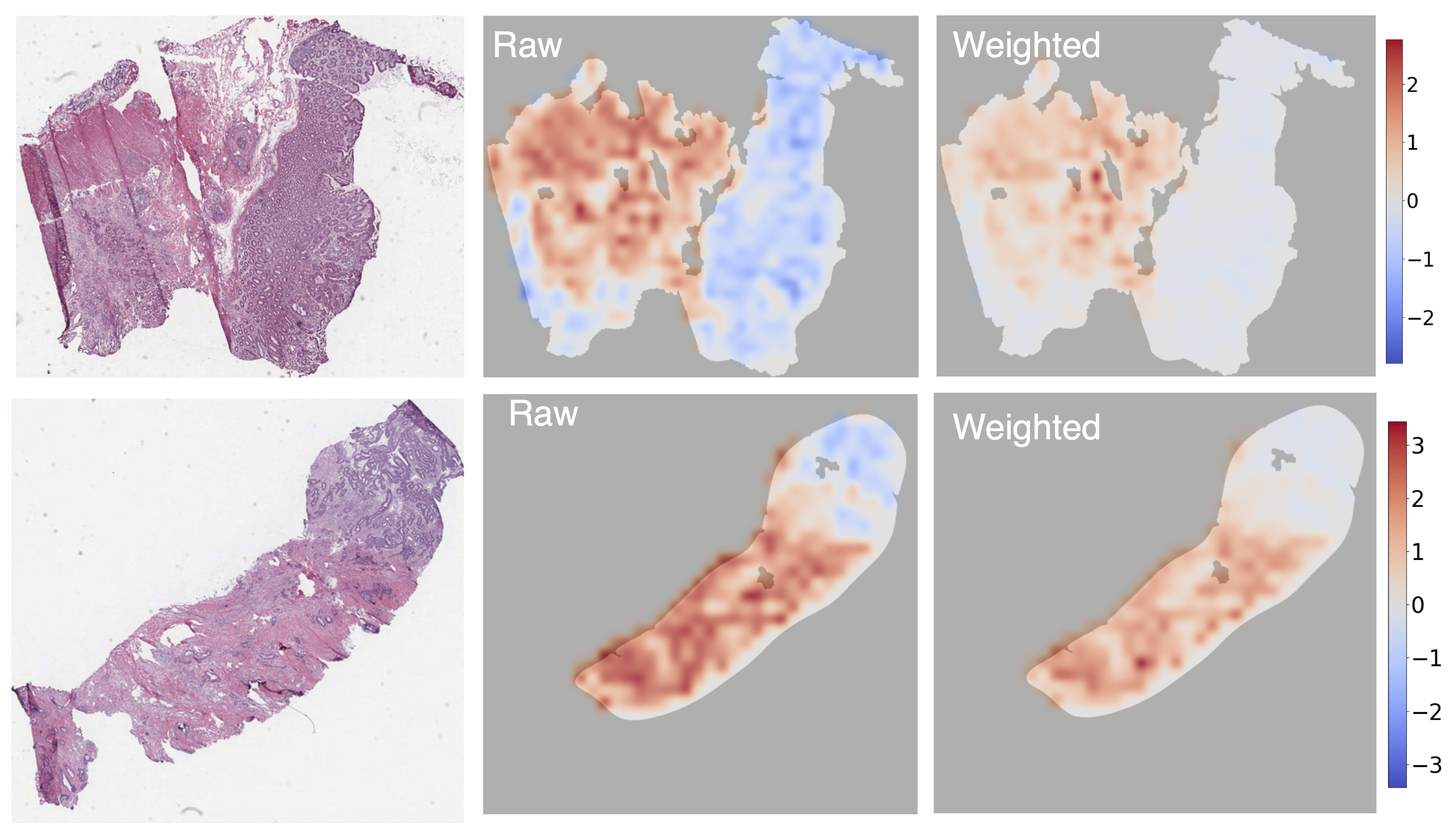}
    \caption{Additional spatialized prediction maps for \textit{AOC3}}
    \label{fig:addpmaoc3}
\end{figure}

\begin{figure}
    \centering
    \includegraphics[width=\textwidth]{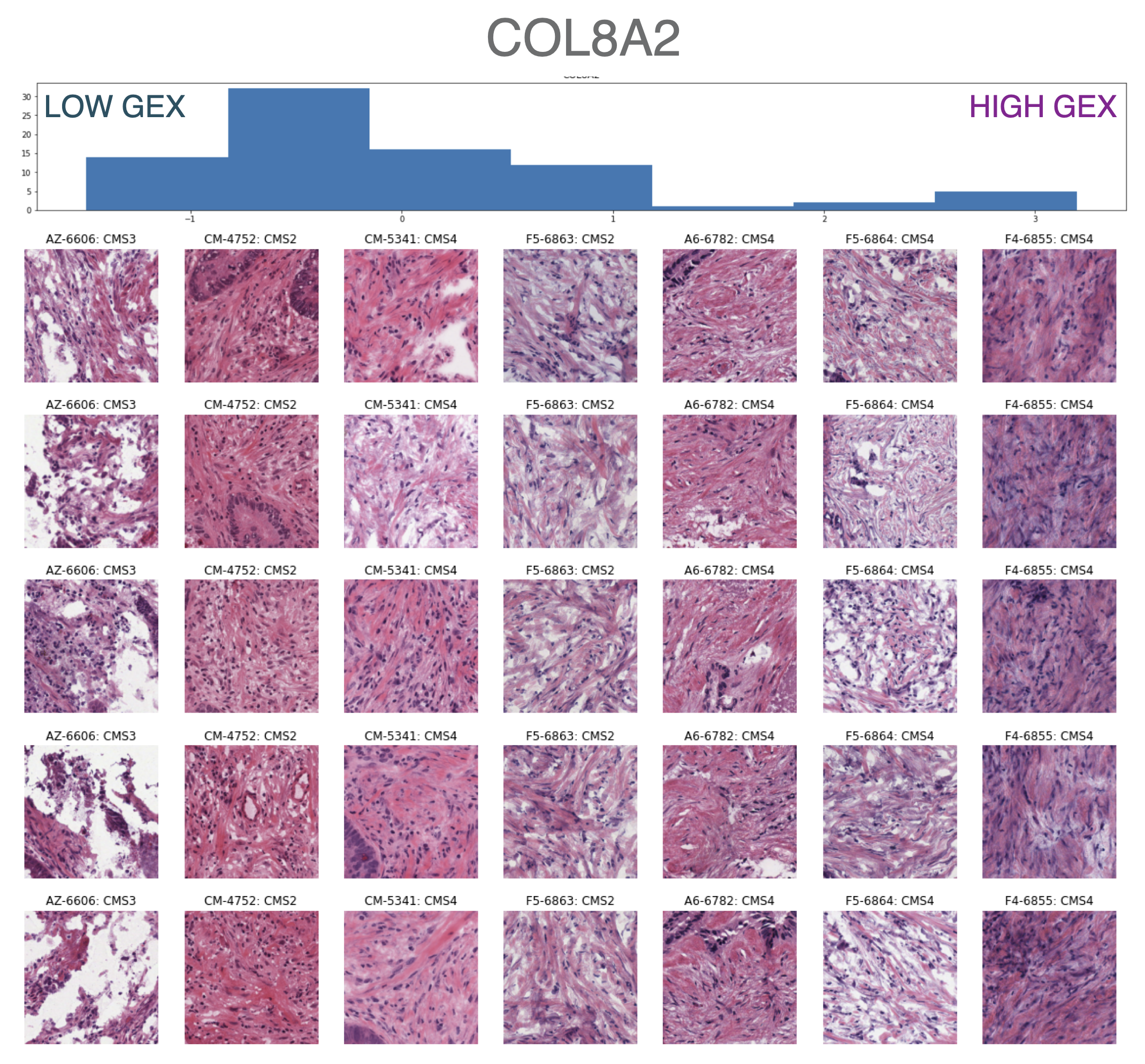}
    \caption{High-attention patches against gene expression levels (gex) for \textit{COL8A2}}
    \label{fig:hiattcol8a2}
\end{figure}

\begin{figure}
    \centering
    \includegraphics[width=\textwidth]{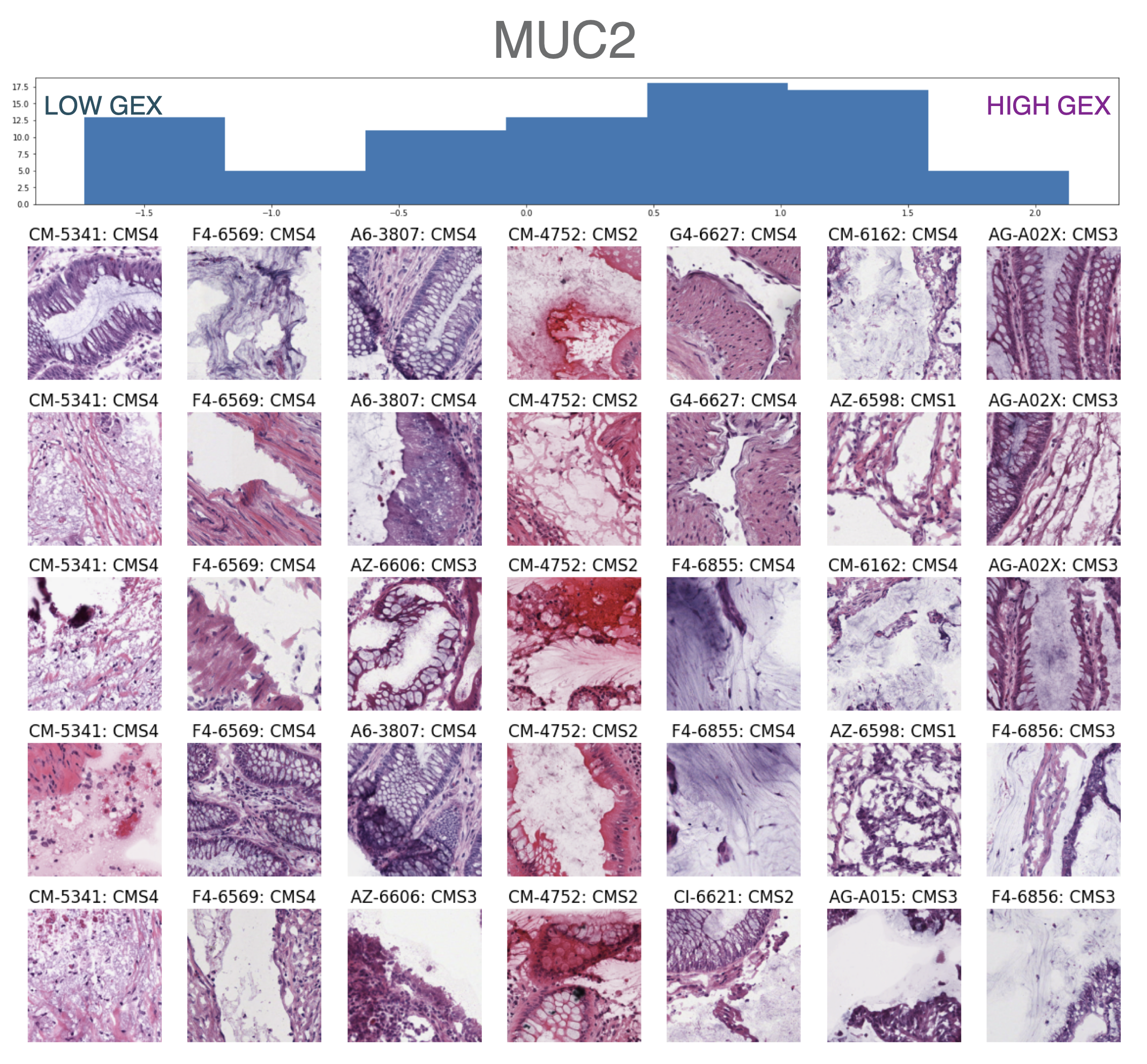}
    \caption{High-attention patches against gene expression levels (gex) for \textit{MUC2}}
    \label{fig:hiattmuc2}
\end{figure}

Additional visualizations of the spatialized prediction maps with and without attention-based weighting are given in Figure~\ref{fig:addpmaoc3}. 
Figures~\ref{fig:hiattcol8a2} and~\ref{fig:hiattmuc2} show in detail the histologic patterns selected by the attention for the genes \textit{COL8A2} and \textit{MUC2} respectively. 

\section{Single-cell co-expression}

\begin{figure}
    \centering
    \begin{subfigure}[b]{0.47\linewidth}
        \includegraphics[width=\linewidth]{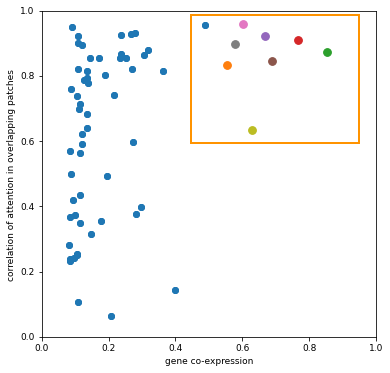}
        \caption{}
    \end{subfigure}
    \hfill
      \begin{subfigure}[b]{0.47\linewidth}
        \includegraphics[width=\linewidth]{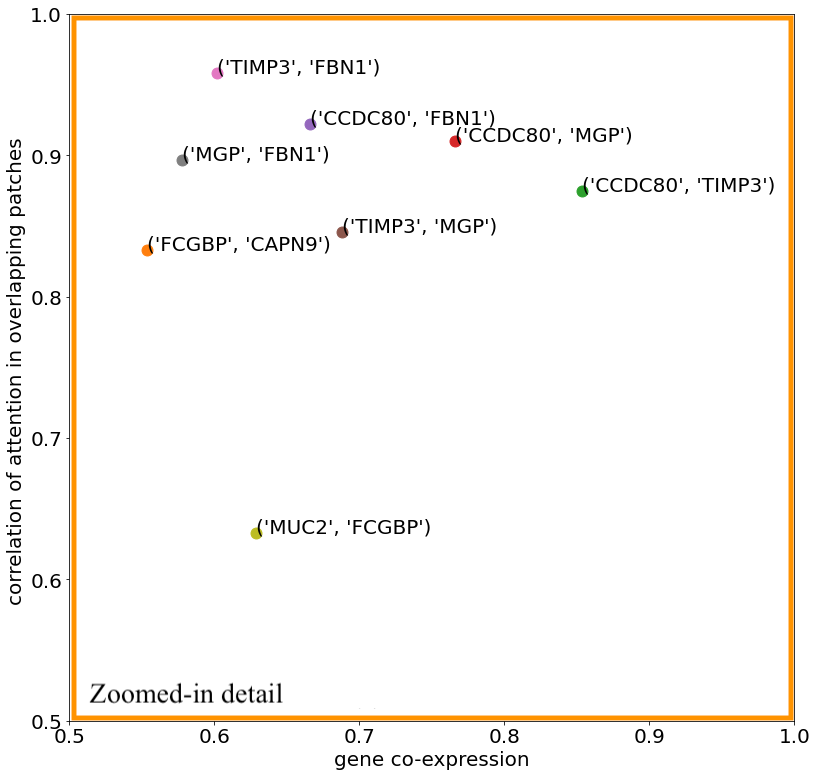}
        \caption{Zoomed in detail from (a)}
    \end{subfigure}
    \caption{Correlation of attention patterns in overlapping WSIs regions for co-expressed genes with correlation p-value$<0.05$.}\label{fig:correlation-analysis}
\end{figure}

\begin{figure}
    \centering
    \includegraphics[width=0.5\textwidth]{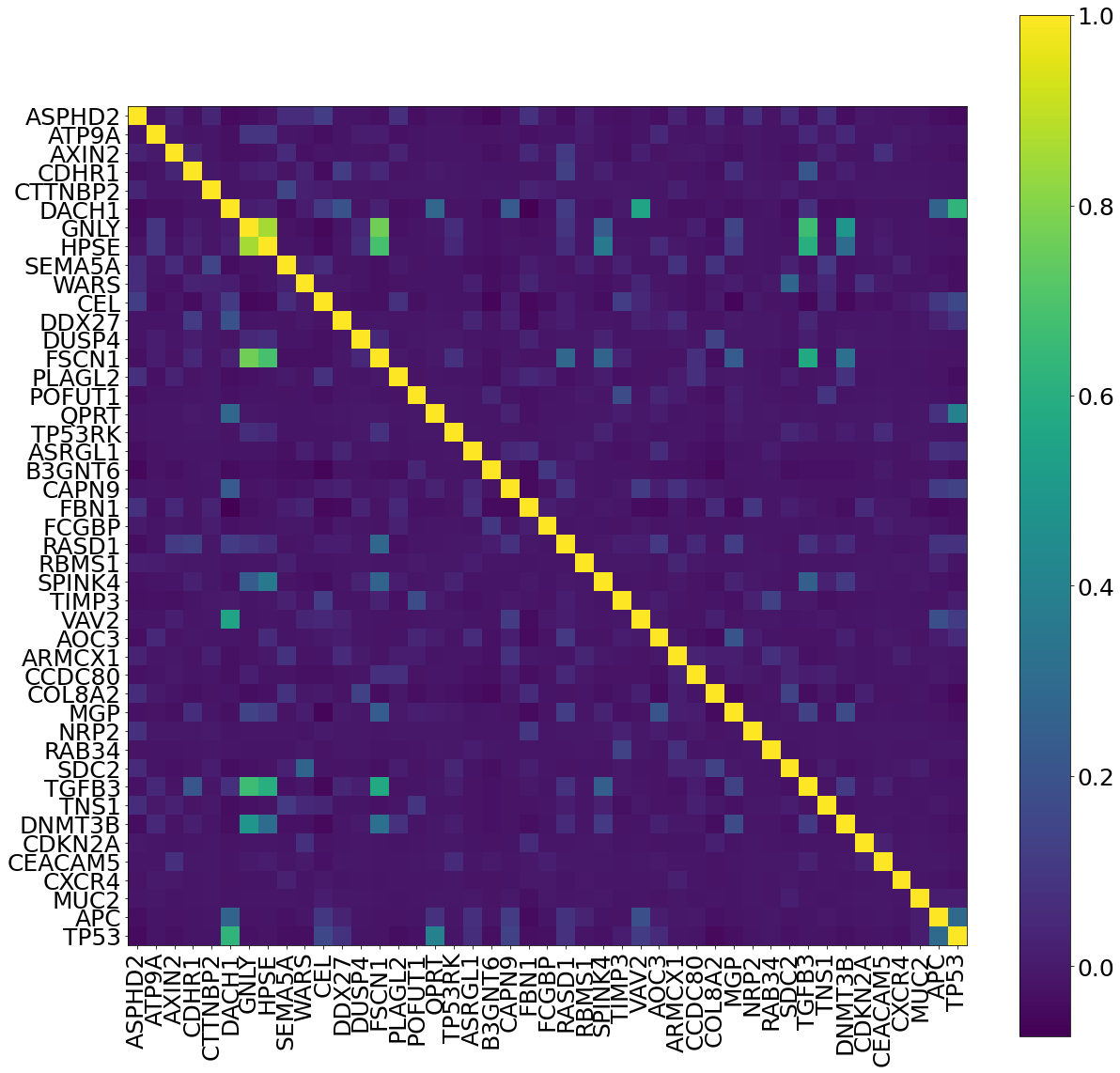}
    \caption{Summary of gene co-expression in single-cells. Each cell in the matrix represents the Pearson's correlation of the gene expressions.}
    \label{fig:scRNA-coexpression}
\end{figure}

The co-expression of genes in single-cell RNA sequencing data is summarized in Figure~\ref{fig:scRNA-coexpression}. 
Figure~\ref{fig:correlation-analysis} compares the correlation of the attention weights in overlapping patches to the co-expression of genes in single-cell data in terms of their Pearson's correlation. 

\begin{figure}
    \centering
    \begin{subfigure}[b]{0.47\linewidth}
        \includegraphics[width=\linewidth]{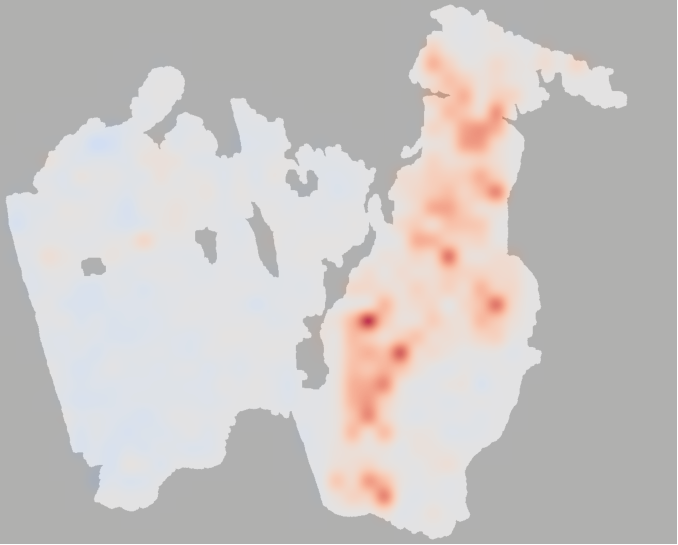}\caption{\textit{FCGBP}}\label{fig:wp-fcgbp}
    \end{subfigure}
    \hfill
      \begin{subfigure}[b]{0.47\linewidth}
        \includegraphics[width=\linewidth]{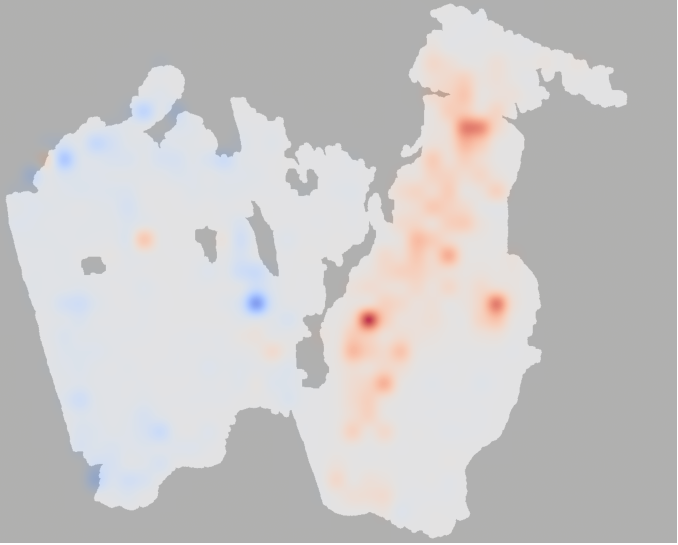}\caption{\textit{MUC2}}\label{fig:wp-muc2-extra}
    \end{subfigure}
    \caption{Attention-weighted predictions for co-expressed genes.}\label{fig:scRNA-attention}
\end{figure}

The couples (i) \textit{CCD80}, \textit{TIMP3}; (ii) \textit{CCDC80}, \textit{MGP}; (iii) \textit{TIMP3}, \textit{MGP}; (iv) \textit{FCGBP}, \textit{MUC2} report high correlation with Pearson's $rho$ being $0.85$, $0.77$, $0.69$, $0.62$ respectively. These couples show high correlation in the attention patterns. 
Figure~\ref{fig:scRNA-attention} shows the similarity in the attention patterns for \textit{FCGBP} and \textit{MUC2}. 

\end{document}